\documentclass[twocolumn]{aastex701}
\newcommand{\target}{KSP-OT-201602a}

\def\arcmin{\hbox{$^\prime$}}

\def\simgt{\lower.5ex\hbox{$\; \buildrel > \over \sim \;$}}%
\def\simlt{\lower.5ex\hbox{$\; \buildrel < \over \sim \;$}}%

\begin{document}

%$BVI$ 
\title{Superoutburst Precursors in the Faint SU UMa-type Dwarf Nova \target}
% Observations of ...
%\title{Photometric Study of a Faint SU UMa-type Dwarf Nova \target}
% KSP-OT-201602a, in K196 Field
% Newly Found
\shorttitle{S. C. Kim et al.}
% K196_1_1_V0050

\correspondingauthor{Sang Chul KIM}
\email{sckim@kasi.re.kr}

\author[0000-0001-9670-1546]{Sang Chul Kim}
\affil{Korea Astronomy and Space Science Institute, 776, Daedeok-daero, Yuseong-gu, Daejeon 34055, Republic of Korea}
\affil{Korea University of Science and Technology (UST), 217 Gajeong-ro, Yuseong-gu, Daejeon 34113, Republic of Korea}
\email{sckim@kasi.re.kr}

\author[0000-0002-6261-1531]{Youngdae Lee}
\affil{Korea Astronomy and Space Science Institute, 776, Daedeok-daero, Yuseong-gu, Daejeon 34055, Republic of Korea}
%\affil{Department of Astronomy and Space Science, Chungnam National %University, Daejeon 34134, Republic of Korea}
\email{youngdaelee@kasi.re.kr}

\author[0000-0003-4200-5064]{Dae-Sik Moon} 
\affil{David A. Dunlap Department of Astronomy and Astrophysics, University of Toronto, 50 St. George Street, Toronto, ON M5S 3H4, Canada}
\email{moon@astro.utoronto.ca}

\author[0000-0002-3505-3036]{Hong Soo Park} 
\affil{Korea Astronomy and Space Science Institute, 776, Daedeok-daero, Yuseong-gu, Daejeon 34055, Republic of Korea}
\affil{Korea University of Science and Technology (UST), 217 Gajeong-ro, Yuseong-gu, Daejeon 34113, Republic of Korea}
\email{hspark@kasi.re.kr}

\author[0000-0001-8748-0016]{Nan Jiang}
\affil{David A. Dunlap Department of Astronomy and Astrophysics, University of Toronto, 50 St. George Street, Toronto, ON M5S 3H4, Canada}
\email{steven.jiang@mail.utoronto.ca}

\shorttitle{Dwarf nova : {\target}}
\shortauthors{S. C. Kim et al.}

%==============================================================================
\begin{abstract}
We present the discovery and photometric study of an SU UMa-type dwarf nova (DN)
  \target.
Korea Microlensing Telescope Network Supernova Program data covered almost 
  four years from 2016 to 2019 in $BVI$-bands for this field, and
  discovered this faint SU UMa-type DN.
Two superoutbursts (SOs: S1 and S2) and two normal outbursts 
  (NOs: N1 and N2) are observed
  of which peak brightnesses are 
  $V=18.95$ mag (S1), $V=18.69$ mag (S2)
  and $V=19.37$ mag (N1), $V=19.11$ mag (N2).
The phase dispersion minimization analysis on superhump of S1
  gives the superhump period of 1.52 hr ($=91.2$ minutes $=0.06333$ days).
  %while the data for S2 are not of sufficient quality to be useful.
The magnitude amplitudes for the SOs are 
  4.34 mag (S1) and 4.60 mag (S2), and those for the NOs are
  3.98 mag (N1) and 4.18 mag (N2).
The durations of the SOs are 11.39 days (S1) and 12.00 days (S2),
  and those for the NOs are 2.68 days (N1) and 3.38 days (N2).
Clear evidence of a precursor is identified in S1 before the SO peak
  and this supports the hypothesis that
  precursor is an NO triggerring an SO and the superhump,
  while a potential precursor based on limited data is only observed in S2.
While the amplitudes of the SOs appear typical for SU UMa-type DNe,
  their durations seem to be quite short 
  which might require further model calculations in the future.
\end{abstract}

\keywords{stars: dwarf novae --- Novae --- Cataclysmic variable stars 
--- Binary Stars --- surveys --- techniques: photometric}

%==============================================================================
\section{Introduction} \label{sec:intro}

Dwarf novae (DNe) are cataclysmic variable stars 
  composed of a white dwarf (WD) primary and 
  a mass-transferring low-mass secondary stars, 
  filling the Roche lobe \citep{warner95,Hellier01}.
  % CVs
The mass transfer from the secondary passing through the Lagrangian point
  accretes matter on the disk of the WD.
DNe are classified into three types according to the morphology of 
  the outburst: SU Ursae Majoris (SU UMa) type with 
  brighter and longer `superoutbursts' (SOs) in addition to normal outbursts (NOs)
  with $\Delta$ mag $\sim0.5-1.0$ mag,
  U Geminorum (U Gem or SS Cygni) type with repetitive outbursts of 
  $2-6$ mag in $V$, and
  Z Camelopardalis (Z Cam) type showing `standstills' 
  staying at intermediate brightness between the peak and quiescent phase
  \citep{kato16,lee24,kim26a}.
SU UMa-types are further classified into subclasses:
  ER UMa \citep{Gao99,Zhao06,lee24}, ordinary SU UMa \citep{kato09Dec,Wood11}, 
  and WZ Sge-types \citep{Kato15WZSge,kim26a},
  in order of decreasing mass transfer rate.

Along with SO, an important characteristic of SU UMa-type systems 
  is the presence of `superhumps`,
  which is a periodic magnitude oscillations
  (amplitude of $\sim 0.1-0.3$ mag; e.g., \citealt{Szegedi22})
  predominantly observed during SOs and
  considered to be from the precessing, tidally distorted
  eccentric accretion disk caused by the 3 : 1 resonance \citep{Whitehurst88,Hirose90,kato09Dec}.
In general, the superhump period ($P_{\rm sh}$) is a few percent 
  longer than the orbital period ($P_{\rm orb}$) 
  \citep{kato16,kim26a}.
In the thermal-tidal instability (TTI) model 
  \citep{Osaki89,Osaki03AA,Osaki13_50},
  the accretion disk radius gradually increases 
  during repetition of `sudden rise-slow decline' patterns
  between successive SOs, %\citep{Osaki13_50}.
  and the SOs are triggered by an NO when the disk radius reaches
  the 3 : 1 resonance.
This NO, during which the 3 : 1 resonance is reached,
  that triggers the SO is called a `precursor' (or precursor outburst)
  lasting one or two days
  \citep{Howell99,Osaki03AA,Osaki05_PJAB,Uemura05,kato16,Kato19_L4}.
These SOs and superhumps arise in systems where 
  the mass of the secondary star is less than $\sim 0.3$ times that of the primary, 
  such that the tidal truncation radius exceeds the radius of the 3:1 resonance
  \citep{Tampo23}.

%\citet{Otula16}

We have been conducting a long-term supernova survey 
  with the KMTNet telescopes \citep{Kim16}, 
  taking advantage of their wide-field coverage and 
  continuous 24-hour monitoring capability.
As a byproduct of this survey, we have identified a large number of DNe
  \citep{Brown18, lee19,lee22,lee24,kim26a,kim26b}.
Some of these systems are located at remarkably large distances 
  \citep{lee19}, suggesting that their donor stars may 
  belong to Population II. 
If confirmed, these systems could represent a distinct 
  evolutionary channel from that of the previous DN population.

In this paper, we present a new SU UMa-type DN discovered by KMTNet
  in 2016 February, which is quite faint 
  with the SO peak only about $V \sim 18.7$ mag (Section~\ref{sec:SOsec}).
This paper is organized as follows.
We describe the discovery and observations in Section~\ref{sec:obs_sec}.
In Section~\ref{sec:result}, we present the main results: 
  light curve analysis, discovery images, SOs, NOs, and 
  the period analysis.
We discuss the precursors in Section~\ref{sec:discuss} and
  summarize our results in Section~\ref{sec:summary}.
    
%==============================================================================
\section{Discovery and Observations}\label{sec:obs_sec}  % Sec 2

Observations are made by using the three 1.6 m 
  KMTNet telescopes located in CTIO (Chile), SSO (Australia), and 
  SAAO (South Africa) \citep{Kim16}.
\target\ is discovered in the KMTNet Supernova Program 
  (KSP; \citealt{Moon16,Chang26}).
The KMTNet telescopes are equipped with wide-field 
  ($2\degr \times 2\degr$ at a pixel scale of 0\farcs4)
  CCD cameras 
  and we conduct imaging observations with 60 s exposures 
  for each of the $BVI$ filters.

The coordinates of \target\ are ($\alpha, \delta$) = 
  ($\rm 13^h 19^m 25.87^s$, $\rm -45\degr 15\arcmin 57\farcs02$) 
  = (199.85779\degr, --45.26584\degr) 
  (J2000)\footnote{$(l,b)$ = (308.09\degr, +17.33\degr)}.
%\ydl{K196 1 1 V0050 RA = 199.8578 DEC = -45.2658}
%\ydl{13h19m25.87s, -45d15m57.02s}
%\ydl{l=308.0857,b= 17.3256}
KMTNet observations of the \target\ field were conducted over a period of 
  3 years and 4 months, from
  MJD $=57434.5794$ (2016 February 16 UT) to 
  MJD $=58648.7511$ (2019 June 14 UT).
The telescope field was centered at irregular galaxy KK196 in Centaurus
  \citep{Karachentsev13}.
\target\ was first discovered at the CTIO on 2016 February 21 
  $05^{h}$ $09^{m}$ $06.47^{s}$ (UT, MJD $=57439.2147$, $B = 19.66 \pm 0.06$ mag),
  $05^{h}$ $11^{m}$ $11.04^{s}$ (UT, MJD $=57439.2161$, $V = 19.45 \pm 0.05$ mag),
  and
  $05^{h}$ $13^{m}$ $26.98^{s}$ (UT, MJD $=57439.2177$, $I = 19.67 \pm 0.05$ 
  mag)\footnote{These magnitudes are standardized values
  after applying $B$-band color correction and $I$-band magnitude shift
  which are explained in the next paragraph, but before the extinction correction.}.
\target\ is not visible in the images previously obtained at SSO
  $\sim 15$ hours ago.

Point-spread function photometry was performed using the DAOPHOT-II package 
  \citep{Ste87},
  and the photometric errors were estimated from the signal-to-noise ratios.
The photometric standardization was made 
  by comparing the measured magnitudes with those of several nearby 
  AAVSO Photometric All Sky Survey\footnote{https://www.aavso.org/apass} 
  % database
  stars contained in the same image as \target.
To standardize the magnitudes, $B$-band color correction was made:
  $\Delta B = B - B' \simeq 0.27 (B-V)$ + \textrm{offset},
  where $B$ and $B'$ are color-corrected $B$-band magnitudes 
  in the standard Johnson system 
  and uncorrected KNTNet $B$-band magnitudes, respectively \citep{Par17,Ni25}.
Since the AAVSO photometry is provided in the $BVi$-bands, 
  an additional correction of $I = i - 0.4$ mag was applied \citep{Par17}.

We obtain the interstellar reddening value of $E(B-V) = 0.095$ mag 
  toward \target, using the Galactic extinction model of \citet{Schlegel98} and 
  the extinction law of \citet{Cardelli89} with $R_V = 3.1$.
Using equation (7) of \citet{Mennickent99},
  the superhump period $P_{\rm sh} = 0.06333$ days of \target\
  (Section~\ref{sec:PDMperiod})
  can be converted to an orbital period of $P_{\rm orb} = 0.06181$ days
  ($=1.48$ hr).
Using equation (4) of \citet{patterson11}, the relation between 
  the SO peak magnitude and orbital period, the absolute magnitude 
  of the SO peak is estimated to be $M_V = 4.66 \pm 0.18$ mag.
Using the peak magnitude of S2 (the brighter SO, see Section~\ref{sec:SOsec}),
  we obtain the extinction-corrected distance modulus of
  $(m-M)_0=V - M_V - A_V= (18.69\pm0.05) - (4.66\pm0.18) - 3.1 \times 0.095 = 
  13.74 \pm 0.23$ mag, and distance of $d = 5.60 \pm 0.60$ kpc.

Using the Galactic coordinates of \target\ and the distance obtained above,
  we can determine the Galactocentric distance 
  $R_{GC} = \sqrt{d^2 + R_0^2 - 2dR_0 \; cos(l) \; cos(b)}
   = 6.7 \pm 0.1$ kpc  % $ = 6.75_{-0.04}^{+0.11}$ kpc
  and the height from the Galactic plane $d \; sin(b) = 
  1.7 \pm 0.2$ kpc,  % 1.69_{-0.17}^{+0.19}$
  assuming $R_0 = 8.3\pm0.2$ kpc for the distance of the Sun from the Galactic center \citep{Reid14}.

In conclusion, the reddening to \target\ is estimated to be 
  $E(B-V) = 0.095$ mag.
The magnitudes given in this study are not extinction corrected values.
Tables~\ref{tab:lcB}, \ref{tab:lcV}, and \ref{tab:lcI} list 
  the $B$, $V$, and $I$-band photometry of {\target}, respectively.

\begin{deluxetable}{ccc ccc ccc ccc}
\tabletypesize{\scriptsize}
\tablecaption{$B$-band photometry of KSP-OT-201602a \label{tab:lcB}}
\tablehead{\colhead{MJD} & \colhead{Mag} & \colhead{Err} & \colhead{MJD} & \colhead{Mag} & \colhead{Err} & \colhead{MJD} & \colhead{Mag} & \colhead{Err} & \colhead{MJD} & \colhead{Mag} & \colhead{Err} \\
\colhead{[days]} & \colhead{[mag]} & \colhead{[mag]} & \colhead{[days]} & \colhead{[mag]} & \colhead{[mag]} & \colhead{[days]} & \colhead{[mag]} & \colhead{[mag]} & \colhead{[days]} & \colhead{[mag]} & \colhead{[mag]}}
\startdata
57439.2147 & 19.66 & 0.06 & 57445.9338 & 19.39 & 0.04 & 57449.9414 & 19.84 & 0.03 & 58194.6342 & 18.88 & 0.02 \\
57439.2325 & 19.55 & 0.05 & 57445.9493 & 19.55 & 0.04 & 57449.9579 & 19.76 & 0.03 & 58194.8883 & 18.99 & 0.02 \\
57439.2467 & 19.53 & 0.05 & 57445.9682 & 19.53 & 0.04 & 57449.9733 & 19.80 & 0.04 & 58194.9553 & 19.07 & 0.03 \\
57439.2606 & 19.52 & 0.04 & 57445.9928 & 19.42 & 0.04 & 57449.9880 & 19.91 & 0.03 & 58195.5650 & 19.26 & 0.02 \\
57439.2749 & 19.41 & 0.04 & 57446.0086 & 19.54 & 0.04 & 57450.0220 & 19.90 & 0.04 & 58195.6278 & 19.29 & 0.02 \\
57439.2898 & 19.51 & 0.06 & 57446.0237 & 19.47 & 0.04 & 57451.1943 & 20.13 & 0.03 & 58195.8736 & 19.28 & 0.02 \\
57439.5846 & 19.49 & 0.09 & 57446.0291 & 19.58 & 0.04 & 57451.2082 & 20.14 & 0.02 & 58195.9434 & 19.33 & 0.02 \\
57439.5997 & 19.75 & 0.15 & 57447.1950 & 19.59 & 0.03 & 57451.2232 & 20.09 & 0.03 & 58196.2096 & 19.30 & 0.01 \\
57439.6144 & 19.40 & 0.12 & 57447.2084 & 19.63 & 0.03 & 57451.2421 & 20.14 & 0.02 & 58196.5757 & 19.39 & 0.02 \\
57439.6291 & 19.52 & 0.12 & 57447.2242 & 19.71 & 0.03 & 57451.2580 & 20.12 & 0.02 & 58196.8843 & 19.47 & 0.02 \\
57439.6443 & 19.60 & 0.12 & 57447.2387 & 19.71 & 0.03 & 57451.5227 & 20.34 & 0.08 & 58196.9520 & 19.45 & 0.02 \\
57439.6593 & 19.53 & 0.09 & 57447.2540 & 19.57 & 0.03 & 57451.5387 & 20.28 & 0.07 & 58197.2052 & 19.61 & 0.02 \\
57439.6747 & 19.35 & 0.09 & 57447.5370 & 19.79 & 0.09 & 57451.5542 & 20.27 & 0.06 & 58198.1990 & 19.71 & 0.03 \\
57441.1664 & 20.66 & 0.13 & 57447.5526 & 19.98 & 0.09 & 57451.5685 & 20.27 & 0.07 & 58199.1743 & 19.79 & 0.02 \\
57441.1902 & 20.72 & 0.16 & 57447.5699 & 19.63 & 0.08 & 57451.5845 & 20.24 & 0.06 & 58200.1666 & 19.83 & 0.02 \\
57441.2040 & 20.80 & 0.15 & 57447.5865 & 19.80 & 0.08 & 57451.6009 & 20.34 & 0.08 & 58200.5872 & 19.80 & 0.03 \\
57441.2310 & 21.22 & 0.20 & 57447.6037 & 19.91 & 0.10 & 57451.6154 & 20.25 & 0.07 & 58200.5890 & 19.88 & 0.04 \\
57441.2480 & 20.97 & 0.15 & 57447.6202 & 19.72 & 0.07 & 57451.6316 & 20.35 & 0.09 & 58202.1266 & 20.26 & 0.04 \\
57441.2625 & 21.16 & 0.15 & 57447.6371 & 19.71 & 0.08 & 57451.6470 & 20.50 & 0.11 & 58202.1926 & 20.22 & 0.02 \\
57441.2771 & 21.28 & 0.17 & 57447.6551 & 19.89 & 0.08 & 57451.9416 & 20.37 & 0.04 & 58202.1946 & 20.22 & 0.02 \\
57441.2910 & 21.26 & 0.17 & 57447.9067 & 19.56 & 0.04 & 57451.9455 & 20.36 & 0.04 & 58202.8668 & 20.73 & 0.13 \\
57441.3052 & 21.16 & 0.15 & 57447.9222 & 19.79 & 0.04 & 57451.9600 & 20.57 & 0.05 & 58202.9311 & 20.64 & 0.10 \\
57443.6050 & 19.30 & 0.05 & 57447.9372 & 19.75 & 0.04 & 57451.9745 & 20.35 & 0.04 & 58555.1275 & 21.24 & 0.08 \\
57443.9205 & 19.31 & 0.04 & 57447.9525 & 19.55 & 0.03 & 57451.9899 & 20.45 & 0.05 & 58555.1895 & 20.79 & 0.05 \\
57443.9357 & 19.35 & 0.05 & 57447.9691 & 19.70 & 0.04 & 57452.0054 & 20.46 & 0.04 & 58555.2397 & 20.59 & 0.04 \\
57443.9510 & 19.10 & 0.04 & 57447.9843 & 19.73 & 0.03 & 57452.0226 & 20.49 & 0.04 & 58555.5274 & 19.59 & 0.03 \\
57443.9682 & 19.18 & 0.05 & 57447.9997 & 19.74 & 0.03 & 57453.1845 & 22.73 & 0.16 & 58555.5761 & 19.61 & 0.03 \\
57443.9908 & 19.28 & 0.05 & 57448.0173 & 19.61 & 0.03 & 57453.2013 & 22.54 & 0.12 & 58555.6272 & 19.64 & 0.04 \\
57444.0114 & 19.18 & 0.05 & 57448.0325 & 19.66 & 0.03 & 57453.2177 & 22.52 & 0.13 & 58555.6315 & 19.65 & 0.03 \\
57445.1458 & 19.38 & 0.04 & 57449.1694 & 19.77 & 0.03 & 57453.2330 & 22.90 & 0.15 & 58555.8928 & 19.74 & 0.04 \\
57445.1601 & 19.26 & 0.04 & 57449.1912 & 19.83 & 0.03 & 57453.2480 & 22.97 & 0.19 & 58557.1918 & 21.67 & 0.09 \\
57445.1738 & 19.36 & 0.04 & 57449.2052 & 19.76 & 0.03 & 57456.1678 & 22.62 & 0.17 & 58557.2389 & 21.69 & 0.08 \\
57445.1876 & 19.40 & 0.03 & 57449.2198 & 19.71 & 0.03 & 57488.4981 & 20.19 & 0.16 & 58557.2430 & 21.83 & 0.09 \\
57445.2011 & 19.34 & 0.03 & 57449.2360 & 19.76 & 0.03 & 57508.8077 & 19.19 & 0.08 & 58574.1226 & 22.01 & 0.19 \\
57445.2170 & 19.23 & 0.03 & 57449.2520 & 19.83 & 0.03 & 57823.9419 & 20.18 & 0.15 & 58641.0163 & 19.47 & 0.02 \\
57445.2317 & 19.34 & 0.04 & 57449.5353 & 19.75 & 0.05 & 57823.9977 & 20.03 & 0.09 & 58641.4080 & 19.41 & 0.08 \\
57445.2476 & 19.36 & 0.03 & 57449.5518 & 19.82 & 0.10 & 58192.1249 & 19.14 & 0.03 & 58642.0171 & 19.76 & 0.03 \\
57445.2620 & 19.41 & 0.03 & 57449.5664 & 20.04 & 0.08 & 58192.1901 & 19.13 & 0.01 & 58642.7534 & 20.62 & 0.06 \\
57445.2783 & 19.30 & 0.03 & 57449.5814 & 20.00 & 0.12 & 58192.5854 & 19.22 & 0.02 & 58643.0196 & 21.22 & 0.09 \\
57445.2936 & 19.28 & 0.03 & 57449.8947 & 19.78 & 0.03 & 58194.1260 & 18.92 & 0.01 &  & &  \\
57445.9009 & 19.47 & 0.05 & 57449.9099 & 19.77 & 0.03 & 58194.1927 & 18.88 & 0.01 &  & &  \\
57445.9175 & 19.36 & 0.04 & 57449.9266 & 19.82 & 0.03 & 58194.5704 & 18.89 & 0.02 &  & & 
\enddata
\tablecomments{165 measurements with error $\leq 0.2$ mag listed in chronological order (top to bottom, then left to right).}
\end{deluxetable}

\begin{deluxetable}{ccc ccc ccc ccc}
\tabletypesize{\scriptsize}
\tablecaption{$V$-band photometry of KSP-OT-201602a \label{tab:lcV}}
\tablehead{\colhead{MJD} & \colhead{Mag} & \colhead{Err} & \colhead{MJD} & \colhead{Mag} & \colhead{Err} & \colhead{MJD} & \colhead{Mag} & \colhead{Err} & \colhead{MJD} & \colhead{Mag} & \colhead{Err} \\
\colhead{[days]} & \colhead{[mag]} & \colhead{[mag]} & \colhead{[days]} & \colhead{[mag]} & \colhead{[mag]} & \colhead{[days]} & \colhead{[mag]} & \colhead{[mag]} & \colhead{[days]} & \colhead{[mag]} & \colhead{[mag]}}
\startdata
57439.2161 & 19.45 & 0.05 & 57445.9356 & 19.21 & 0.04 & 57449.9604 & 19.71 & 0.04 & 58192.1916 & 18.95 & 0.02 \\
57439.2340 & 19.34 & 0.04 & 57445.9509 & 19.25 & 0.04 & 57449.9749 & 19.58 & 0.04 & 58192.5868 & 19.02 & 0.02 \\
57439.2482 & 19.28 & 0.04 & 57445.9699 & 19.31 & 0.04 & 57449.9895 & 19.63 & 0.04 & 58194.1274 & 18.69 & 0.02 \\
57439.2621 & 19.28 & 0.04 & 57445.9945 & 19.18 & 0.04 & 57450.0058 & 19.59 & 0.04 & 58194.1942 & 18.73 & 0.01 \\
57439.2764 & 19.27 & 0.04 & 57446.0105 & 19.34 & 0.04 & 57451.1958 & 19.83 & 0.03 & 58194.5717 & 18.66 & 0.02 \\
57439.2913 & 19.27 & 0.05 & 57446.0254 & 19.30 & 0.05 & 57451.2097 & 19.88 & 0.03 & 58194.6355 & 18.70 & 0.02 \\
57439.5864 & 19.15 & 0.08 & 57446.0308 & 19.31 & 0.04 & 57451.2250 & 19.82 & 0.03 & 58194.8898 & 18.77 & 0.02 \\
57439.6015 & 19.32 & 0.09 & 57447.1965 & 19.43 & 0.03 & 57451.2437 & 19.88 & 0.03 & 58194.9568 & 18.91 & 0.02 \\
57439.6160 & 19.30 & 0.10 & 57447.2100 & 19.51 & 0.04 & 57451.2595 & 19.94 & 0.03 & 58195.5664 & 19.05 & 0.02 \\
57439.6309 & 19.24 & 0.10 & 57447.2257 & 19.51 & 0.04 & 57451.5242 & 20.07 & 0.08 & 58195.6293 & 19.09 & 0.02 \\
57439.6459 & 19.18 & 0.07 & 57447.2402 & 19.50 & 0.04 & 57451.5406 & 20.07 & 0.07 & 58195.8750 & 19.15 & 0.02 \\
57439.6609 & 19.15 & 0.06 & 57447.2556 & 19.41 & 0.03 & 57451.5557 & 19.93 & 0.05 & 58195.9449 & 19.16 & 0.02 \\
57439.6773 & 19.07 & 0.07 & 57447.5391 & 19.70 & 0.09 & 57451.5702 & 20.09 & 0.07 & 58196.2109 & 19.04 & 0.02 \\
57441.1680 & 20.65 & 0.15 & 57447.5544 & 19.56 & 0.08 & 57451.5861 & 19.98 & 0.06 & 58196.5771 & 19.23 & 0.03 \\
57441.1774 & 20.70 & 0.15 & 57447.5715 & 19.50 & 0.08 & 57451.6026 & 20.06 & 0.07 & 58196.8857 & 19.23 & 0.02 \\
57441.1917 & 20.42 & 0.13 & 57447.5891 & 19.60 & 0.07 & 57451.6173 & 19.95 & 0.07 & 58196.9535 & 19.30 & 0.02 \\
57441.2054 & 20.36 & 0.10 & 57447.6052 & 19.47 & 0.06 & 57451.6332 & 20.06 & 0.08 & 58197.2065 & 19.35 & 0.02 \\
57441.2191 & 20.61 & 0.15 & 57447.6217 & 19.69 & 0.10 & 57451.6486 & 20.07 & 0.09 & 58198.2005 & 19.52 & 0.03 \\
57441.2325 & 20.94 & 0.19 & 57447.6388 & 19.55 & 0.08 & 57451.9432 & 20.12 & 0.04 & 58199.1756 & 19.54 & 0.02 \\
57441.2642 & 20.64 & 0.13 & 57447.6568 & 19.53 & 0.06 & 57451.9470 & 20.10 & 0.04 & 58200.1680 & 19.60 & 0.02 \\
57441.2785 & 20.73 & 0.12 & 57447.9083 & 19.44 & 0.04 & 57451.9616 & 20.22 & 0.04 & 58200.5904 & 19.63 & 0.04 \\
57441.2927 & 21.04 & 0.16 & 57447.9238 & 19.51 & 0.04 & 57451.9761 & 20.14 & 0.05 & 58202.1281 & 19.97 & 0.04 \\
57441.3068 & 20.88 & 0.11 & 57447.9389 & 19.56 & 0.04 & 57451.9920 & 20.11 & 0.05 & 58202.1960 & 19.95 & 0.03 \\
57443.6066 & 19.09 & 0.05 & 57447.9544 & 19.36 & 0.03 & 57452.0069 & 20.22 & 0.04 & 58202.8682 & 20.41 & 0.12 \\
57443.9222 & 19.05 & 0.04 & 57447.9708 & 19.41 & 0.03 & 57452.0245 & 20.22 & 0.04 & 58202.9326 & 20.45 & 0.10 \\
57443.9374 & 19.09 & 0.05 & 57447.9860 & 19.51 & 0.04 & 57453.1271 & 22.20 & 0.20 & 58223.8071 & 14.02 & 0.02 \\
57443.9527 & 18.83 & 0.04 & 57448.0014 & 19.48 & 0.04 & 57453.1716 & 22.34 & 0.17 & 58553.1240 & 21.02 & 0.10 \\
57443.9699 & 18.99 & 0.04 & 57448.0190 & 19.31 & 0.03 & 57453.2029 & 22.40 & 0.14 & 58555.1288 & 20.81 & 0.09 \\
57443.9926 & 19.05 & 0.05 & 57448.0342 & 19.45 & 0.03 & 57453.2193 & 22.35 & 0.13 & 58555.1910 & 20.55 & 0.06 \\
57444.0131 & 18.93 & 0.05 & 57449.1709 & 19.56 & 0.03 & 57453.2346 & 22.33 & 0.14 & 58555.2410 & 20.28 & 0.05 \\
57444.0306 & 18.92 & 0.04 & 57449.1880 & 19.60 & 0.03 & 57453.2495 & 22.50 & 0.19 & 58555.5289 & 19.38 & 0.03 \\
57445.1472 & 19.18 & 0.04 & 57449.1932 & 19.61 & 0.04 & 57456.1857 & 22.37 & 0.19 & 58555.5774 & 19.46 & 0.03 \\
57445.1616 & 19.12 & 0.03 & 57449.2067 & 19.55 & 0.04 & 57456.2144 & 22.51 & 0.20 & 58555.6285 & 19.38 & 0.03 \\
57445.1753 & 19.17 & 0.04 & 57449.2212 & 19.54 & 0.03 & 57488.5139 & 20.54 & 0.15 & 58557.1932 & 21.51 & 0.09 \\
57445.1891 & 19.15 & 0.04 & 57449.2382 & 19.55 & 0.03 & 57489.4683 & 21.12 & 0.18 & 58557.2402 & 21.69 & 0.10 \\
57445.2026 & 19.21 & 0.03 & 57449.2536 & 19.62 & 0.03 & 57489.5101 & 20.09 & 0.19 & 58573.1882 & 20.94 & 0.11 \\
57445.2184 & 18.99 & 0.03 & 57449.5368 & 19.48 & 0.04 & 57489.5379 & 19.95 & 0.17 & 58573.9472 & 20.80 & 0.20 \\
57445.2332 & 19.09 & 0.03 & 57449.5534 & 19.49 & 0.05 & 57512.4688 & 20.85 & 0.17 & 58641.0176 & 19.26 & 0.03 \\
57445.2494 & 19.25 & 0.04 & 57449.5680 & 19.74 & 0.07 & 57517.0703 & 20.48 & 0.20 & 58641.4095 & 19.13 & 0.09 \\
57445.2635 & 19.25 & 0.03 & 57449.5829 & 19.57 & 0.12 & 57823.9434 & 20.32 & 0.15 & 58641.7561 & 19.31 & 0.05 \\
57445.2798 & 19.06 & 0.03 & 57449.8963 & 19.71 & 0.03 & 57823.9992 & 19.68 & 0.10 & 58642.0184 & 19.54 & 0.03 \\
57445.2951 & 19.12 & 0.03 & 57449.9115 & 19.50 & 0.03 & 57824.0023 & 19.79 & 0.09 & 58642.7583 & 20.47 & 0.06 \\
57445.9027 & 19.30 & 0.05 & 57449.9283 & 19.48 & 0.12 & 57843.9194 & 21.15 & 0.16 & 58643.0210 & 21.06 & 0.11 \\
57445.9193 & 19.10 & 0.04 & 57449.9431 & 19.66 & 0.04 & 58192.1263 & 18.95 & 0.02 &  & & 
\enddata
\tablecomments{175 measurements with error $\leq 0.2$ mag listed in chronological order (top to bottom, then left to right).}
\end{deluxetable}

\begin{deluxetable}{ccc ccc ccc ccc}
\tabletypesize{\scriptsize}
\tablecaption{$I$-band photometry of KSP-OT-201602a \label{tab:lcI}}
\tablehead{\colhead{MJD} & \colhead{Mag} & \colhead{Err} & \colhead{MJD} & \colhead{Mag} & \colhead{Err} & \colhead{MJD} & \colhead{Mag} & \colhead{Err} & \colhead{MJD} & \colhead{Mag} & \colhead{Err} \\
\colhead{[days]} & \colhead{[mag]} & \colhead{[mag]} & \colhead{[days]} & \colhead{[mag]} & \colhead{[mag]} & \colhead{[days]} & \colhead{[mag]} & \colhead{[mag]} & \colhead{[days]} & \colhead{[mag]} & \colhead{[mag]}}
\startdata
57439.2177 & 19.67 & 0.05 & 57445.9373 & 19.52 & 0.05 & 57451.5573 & 20.10 & 0.07 & 58194.8913 & 19.12 & 0.03 \\
57439.2357 & 19.68 & 0.05 & 57445.9551 & 19.62 & 0.04 & 57451.5718 & 20.29 & 0.10 & 58194.9583 & 19.20 & 0.04 \\
57439.2497 & 19.56 & 0.05 & 57445.9718 & 19.55 & 0.05 & 57451.5886 & 20.36 & 0.10 & 58195.5679 & 19.37 & 0.03 \\
57439.2636 & 19.53 & 0.05 & 57445.9962 & 19.55 & 0.06 & 57451.6042 & 20.41 & 0.12 & 58195.6307 & 19.38 & 0.03 \\
57439.2780 & 19.61 & 0.05 & 57446.0122 & 19.61 & 0.05 & 57451.6191 & 20.33 & 0.11 & 58195.8763 & 19.49 & 0.04 \\
57439.2928 & 19.53 & 0.06 & 57446.0272 & 19.60 & 0.05 & 57451.6349 & 20.52 & 0.16 & 58195.9464 & 19.49 & 0.03 \\
57439.5880 & 19.36 & 0.08 & 57446.0325 & 19.65 & 0.05 & 57451.6518 & 20.28 & 0.12 & 58196.2123 & 19.32 & 0.03 \\
57439.6031 & 19.42 & 0.09 & 57447.1980 & 19.72 & 0.04 & 57451.9485 & 20.35 & 0.05 & 58196.5784 & 19.46 & 0.04 \\
57439.6176 & 19.37 & 0.08 & 57447.2115 & 19.77 & 0.05 & 57451.9636 & 20.46 & 0.07 & 58196.8872 & 19.57 & 0.04 \\
57439.6324 & 19.49 & 0.08 & 57447.2272 & 19.84 & 0.06 & 57451.9778 & 20.38 & 0.06 & 58196.9550 & 19.54 & 0.03 \\
57439.6475 & 19.38 & 0.07 & 57447.2418 & 19.85 & 0.06 & 57451.9936 & 20.33 & 0.05 & 58197.2080 & 19.62 & 0.03 \\
57439.6790 & 19.49 & 0.08 & 57447.2570 & 19.69 & 0.04 & 57452.0084 & 20.42 & 0.06 & 58198.2018 & 19.77 & 0.04 \\
57441.1695 & 21.20 & 0.17 & 57447.5407 & 20.07 & 0.13 & 57452.0276 & 20.45 & 0.06 & 58199.1770 & 19.83 & 0.03 \\
57441.1788 & 21.25 & 0.18 & 57447.5560 & 19.88 & 0.12 & 57453.1586 & 21.83 & 0.18 & 58200.1695 & 19.97 & 0.03 \\
57441.1932 & 20.90 & 0.13 & 57447.5732 & 19.88 & 0.10 & 57453.1894 & 22.24 & 0.16 & 58200.5920 & 19.80 & 0.05 \\
57441.2070 & 20.99 & 0.13 & 57447.5907 & 19.90 & 0.11 & 57453.2046 & 22.32 & 0.19 & 58202.1296 & 20.28 & 0.06 \\
57441.2205 & 21.30 & 0.20 & 57447.6069 & 19.86 & 0.12 & 57453.2208 & 22.34 & 0.15 & 58202.1977 & 20.26 & 0.05 \\
57441.2341 & 21.21 & 0.13 & 57447.6234 & 19.78 & 0.09 & 57453.2361 & 22.18 & 0.15 & 58202.8696 & 21.09 & 0.17 \\
57441.2512 & 20.80 & 0.11 & 57447.6403 & 19.65 & 0.06 & 57460.9737 & 19.65 & 0.08 & 58202.9341 & 20.57 & 0.10 \\
57441.2658 & 20.96 & 0.12 & 57447.6584 & 19.74 & 0.08 & 57466.2310 & 21.80 & 0.18 & 58226.0664 & 19.78 & 0.08 \\
57441.2803 & 21.11 & 0.13 & 57447.9102 & 19.80 & 0.05 & 57468.9054 & 20.07 & 0.12 & 58235.7858 & 19.85 & 0.18 \\
57441.2942 & 21.15 & 0.13 & 57447.9255 & 19.73 & 0.05 & 57468.9484 & 19.93 & 0.10 & 58237.7845 & 19.54 & 0.10 \\
57441.3085 & 21.35 & 0.15 & 57447.9405 & 19.88 & 0.06 & 57468.9635 & 20.00 & 0.13 & 58237.8515 & 19.81 & 0.11 \\
57441.6700 & 21.01 & 0.20 & 57447.9562 & 19.58 & 0.04 & 57489.5115 & 19.78 & 0.15 & 58249.7828 & 20.09 & 0.10 \\
57441.6754 & 21.17 & 0.19 & 57447.9725 & 19.76 & 0.04 & 57489.5397 & 19.71 & 0.10 & 58252.4364 & 19.92 & 0.08 \\
57441.9570 & 21.56 & 0.20 & 57447.9877 & 19.77 & 0.05 & 57504.0650 & 19.80 & 0.13 & 58269.4214 & 19.08 & 0.18 \\
57441.9724 & 21.20 & 0.18 & 57448.0030 & 19.84 & 0.05 & 57504.0791 & 19.50 & 0.11 & 58548.5724 & 19.77 & 0.19 \\
57442.0353 & 21.62 & 0.16 & 57448.0207 & 19.65 & 0.05 & 57504.1053 & 19.92 & 0.10 & 58548.6219 & 19.67 & 0.11 \\
57443.6083 & 19.41 & 0.06 & 57448.0359 & 19.75 & 0.04 & 57519.4571 & 20.01 & 0.14 & 58555.1302 & 21.28 & 0.16 \\
57443.6230 & 19.54 & 0.07 & 57449.1726 & 19.85 & 0.05 & 57533.7913 & 19.69 & 0.07 & 58555.1923 & 20.76 & 0.09 \\
57443.6525 & 19.20 & 0.04 & 57449.2227 & 19.95 & 0.06 & 57802.6787 & 19.77 & 0.14 & 58555.2424 & 20.63 & 0.08 \\
57443.9238 & 19.33 & 0.05 & 57449.2405 & 19.89 & 0.05 & 57802.7324 & 19.72 & 0.12 & 58555.5304 & 19.70 & 0.04 \\
57443.9391 & 19.40 & 0.06 & 57449.2552 & 20.02 & 0.06 & 57803.7748 & 19.86 & 0.15 & 58555.6300 & 19.74 & 0.04 \\
57443.9543 & 19.08 & 0.04 & 57449.5384 & 19.82 & 0.07 & 57813.9748 & 19.88 & 0.06 & 58557.1945 & 21.96 & 0.18 \\
57443.9715 & 19.31 & 0.05 & 57449.5550 & 20.00 & 0.17 & 57821.9432 & 19.65 & 0.11 & 58559.9837 & 19.85 & 0.08 \\
57443.9943 & 19.49 & 0.05 & 57449.5696 & 19.86 & 0.09 & 57823.9449 & 20.43 & 0.10 & 58563.9533 & 20.00 & 0.13 \\
57444.0149 & 19.18 & 0.05 & 57449.8978 & 19.88 & 0.04 & 57824.0007 & 20.04 & 0.07 & 58573.9488 & 19.80 & 0.09 \\
57444.0173 & 19.15 & 0.04 & 57449.9130 & 19.84 & 0.05 & 57830.9316 & 19.63 & 0.08 & 58575.9116 & 19.73 & 0.13 \\
57444.0323 & 19.32 & 0.05 & 57449.9298 & 19.93 & 0.05 & 57870.4542 & 19.91 & 0.18 & 58577.1250 & 19.65 & 0.04 \\
57445.1488 & 19.48 & 0.05 & 57449.9450 & 20.00 & 0.05 & 57875.4545 & 19.42 & 0.11 & 58600.5222 & 19.90 & 0.12 \\
57445.1631 & 19.43 & 0.04 & 57449.9620 & 19.97 & 0.06 & 57886.8025 & 19.88 & 0.09 & 58603.8255 & 19.65 & 0.06 \\
57445.1768 & 19.39 & 0.05 & 57449.9764 & 19.73 & 0.05 & 57902.4063 & 20.11 & 0.12 & 58605.8236 & 19.86 & 0.05 \\
57445.1906 & 19.57 & 0.05 & 57449.9911 & 19.89 & 0.05 & 57909.4004 & 19.77 & 0.15 & 58608.4819 & 19.87 & 0.06 \\
57445.2041 & 19.59 & 0.05 & 57450.0074 & 19.99 & 0.05 & 57913.7647 & 16.18 & 0.20 & 58614.4454 & 19.86 & 0.11 \\
57445.2209 & 19.41 & 0.04 & 57451.1974 & 20.23 & 0.06 & 57925.3465 & 19.59 & 0.05 & 58625.0394 & 19.76 & 0.06 \\
57445.2347 & 19.47 & 0.05 & 57451.2112 & 20.18 & 0.05 & 58192.1276 & 19.28 & 0.03 & 58641.0189 & 19.69 & 0.06 \\
57445.2512 & 19.60 & 0.05 & 57451.2266 & 20.11 & 0.05 & 58192.1931 & 19.25 & 0.03 & 58642.0198 & 19.84 & 0.05 \\
57445.2652 & 19.57 & 0.04 & 57451.2452 & 20.22 & 0.05 & 58192.5882 & 19.33 & 0.03 & 58642.7598 & 20.78 & 0.09 \\
57445.2814 & 19.34 & 0.04 & 57451.2486 & 20.17 & 0.05 & 58194.1289 & 19.06 & 0.02 & 58643.0223 & 21.50 & 0.18 \\
57445.2968 & 19.45 & 0.05 & 57451.2611 & 20.16 & 0.04 & 58194.1956 & 19.04 & 0.02 &  & &  \\
57445.9044 & 19.53 & 0.05 & 57451.5259 & 20.29 & 0.11 & 58194.5732 & 19.02 & 0.02 &  & &  \\
57445.9211 & 19.44 & 0.04 & 57451.5424 & 20.38 & 0.10 & 58194.6370 & 19.04 & 0.02 &  & & 
\enddata
\tablecomments{205 measurements with error $\leq 0.2$ mag listed in chronological order (top to bottom, then left to right).}
\end{deluxetable}

%==============================================================================
\section{Results} \label{sec:result}  % Sec 3

\subsection{Light Curve Analysis}    \label{sec:lightcurve}
\begin{figure*}   % Fig 1 ---------------------------
\epsscale{1.1}
\plotone{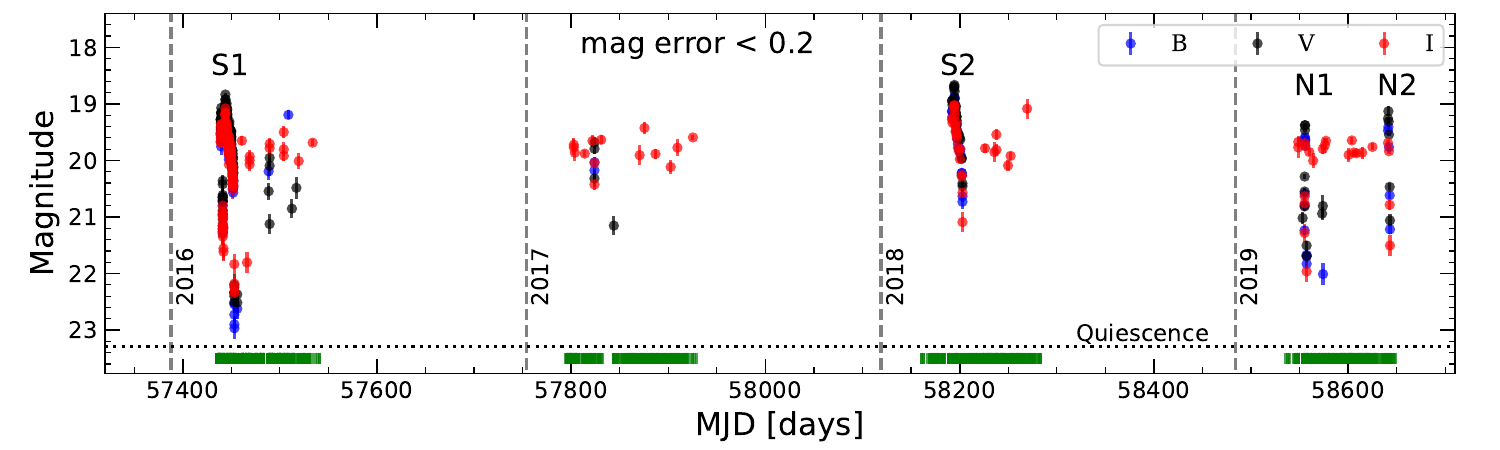}
\caption{$BVI$-band light curve of \target, with $B$-band in blue,
  $V$ in black, and $I$ in red circles.
The KMTNet observations covered almost four years from 2016 to 2019,
  and gray vertical dashed lines show the start of the years.
Two superoutbursts (S1 and S2) and two normal outbursts (N1 and N2)
  are labeled.
Green vertical lines at the bottom show the epochs 
  when the KMTNet observations in $V$-band are made and
  the horizontal black dotted line shows the quiescent phase magnitude
  obtained from 30 stacked $V$-band image ($V\approx 23.29 \pm 0.17$ mag).
}
\label{fig:lcfull}
\end{figure*}

Figure~\ref{fig:lcfull} shows the full light curve of \target\
  from 2016 to 2019 for almost four years 
  with three solar conjunction periods in between and 
  with magnitude errors $< 0.2$ mag.
Two SOs are seen in 2016 (S1) and 2018 (S2), and
  two NOs are observed in 2019 (N1 and N2).
More analyses on these outbursts are in Section~\ref{sec:SOsec} (for SOs)
  and in Section~\ref{sec:NOsec} (for NOs).
Having both of the SOs and NOs together with the superhump period
  $P_{\rm sh} \approx 1.52$ hr (Section~\ref{sec:PDMperiod})
  implying similar orbital period demonstrates that
  \target\ is an SU UMa-type DN \citep[e.g.,][]{kato09Dec}.

To estimate the magnitudes during the quiescent phase, we stacked 30 images
  in each of the $BVI$-bands, resulting in a total exposure time of $30 \times 60$ seconds
  per each band.
This procedure was repeated five times 
  by selecting 30 images from various epochs,
  and we finally obtained the quiescent phase
  magnitudes by taking average of the five estimates:
  $B \approx 23.67 \pm 0.19$, $V \approx 23.29 \pm 0.17$, and 
  $I \approx 23.11 \pm 0.21$ mag.
From these values, we obtain the quiescent phase colors of 
  $B-V \approx 0.38 \pm 0.25$, $V-I \approx 0.17 \pm 0.27$, and
  $B-I \approx 0.56 \pm 0.28$ mag.

%--------------------------------------------------------------------------------
\subsection{Images}         \label{sec:Imag}

\begin{figure*}   % Fig 2 ---------------------------
\epsscale{1.0} %{0.8}
\plotone{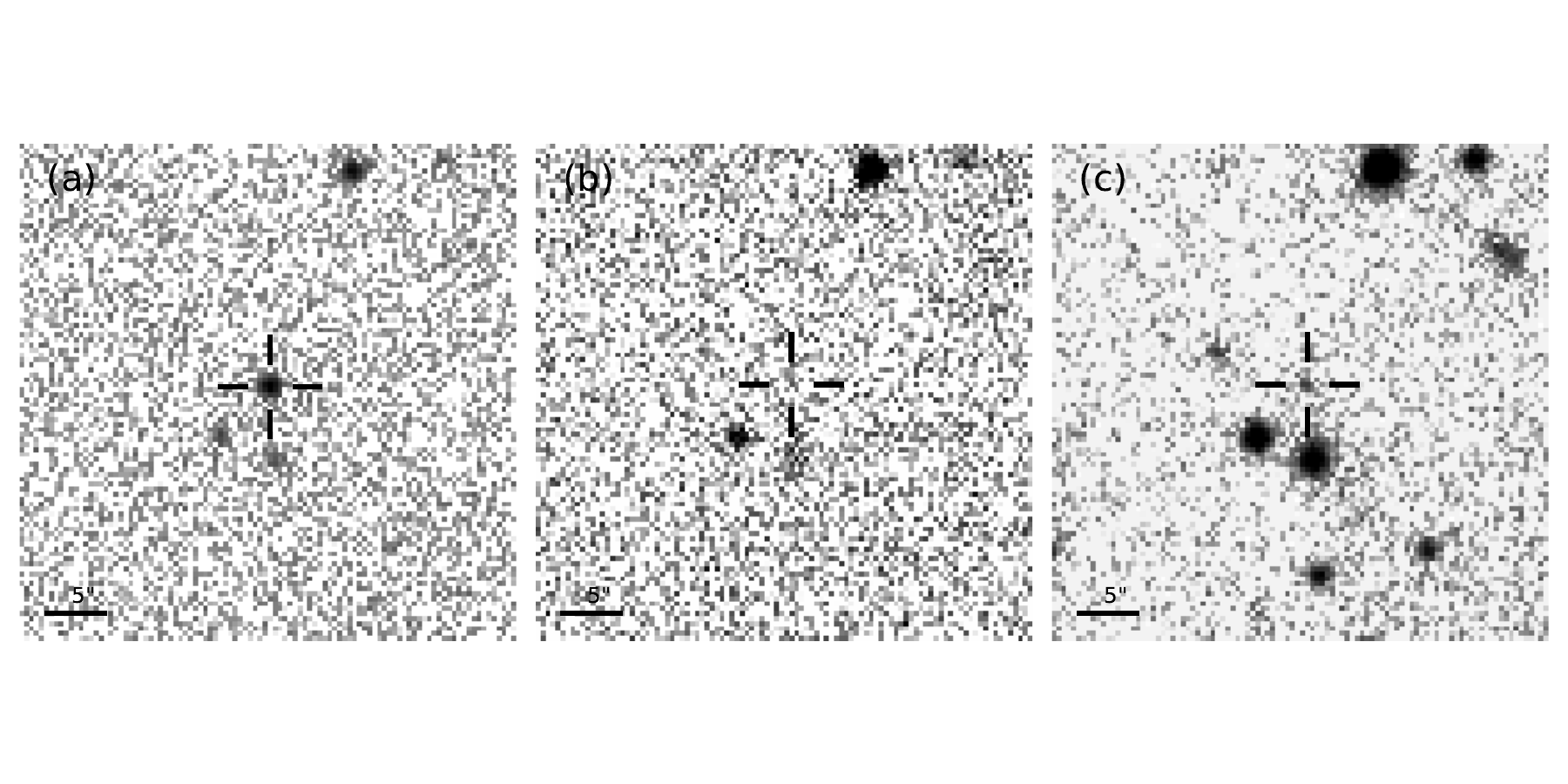}
\caption{(a) $V$-band single image (exposure $=60$ sec) near the peak brightness of S1, 
  (b) another single image at $\sim 1$ month after the peak and
  (c) 30 stacked $V$-band image (exposure $\sim 30 \times 60$ sec)
  at the quiescent phase resulting in $V \approx 23.29 \pm 0.17$ mag.
}
\label{fig:imagthree}
\end{figure*}

Figure~\ref{fig:imagthree} shows (a) a single $V$-band image 
  (exposure $=60$ sec) near the peak of S1
  at 2016 February 27 $05^{h}$ $59^{m}$ (UT, MJD$=57445.24931$),
  % 	57445.24930556, Chile
  (b) another single $V$-band image at $\sim 1$ month after the peak
  at 2016 March 29 $02^{h}$ $17^{m}$ (UT, MJD$=57476.09514$)
  representing the quiescent phase,
  % 57476.09513889
  and (c) 30 stacked image (exposure $\sim 30 \times 60$ sec)
  at quiescent phase from 2016 February 17 to 2016 February 19 (UT,
  MJD$=57435.21334 - 57437.23534$).
  % 2016 February 17 05:07:12.720 UT - 2016 February 19 05:38:53.800 UT
  % MJD$=57435.21334166667 - 57437.23534490741$

%--------------------------------------------------------------------------------
\subsection{Superoutbursts}         \label{sec:SOsec}

\begin{figure*}   % Fig 3 ---------------------------
\epsscale{1.0}
\plotone{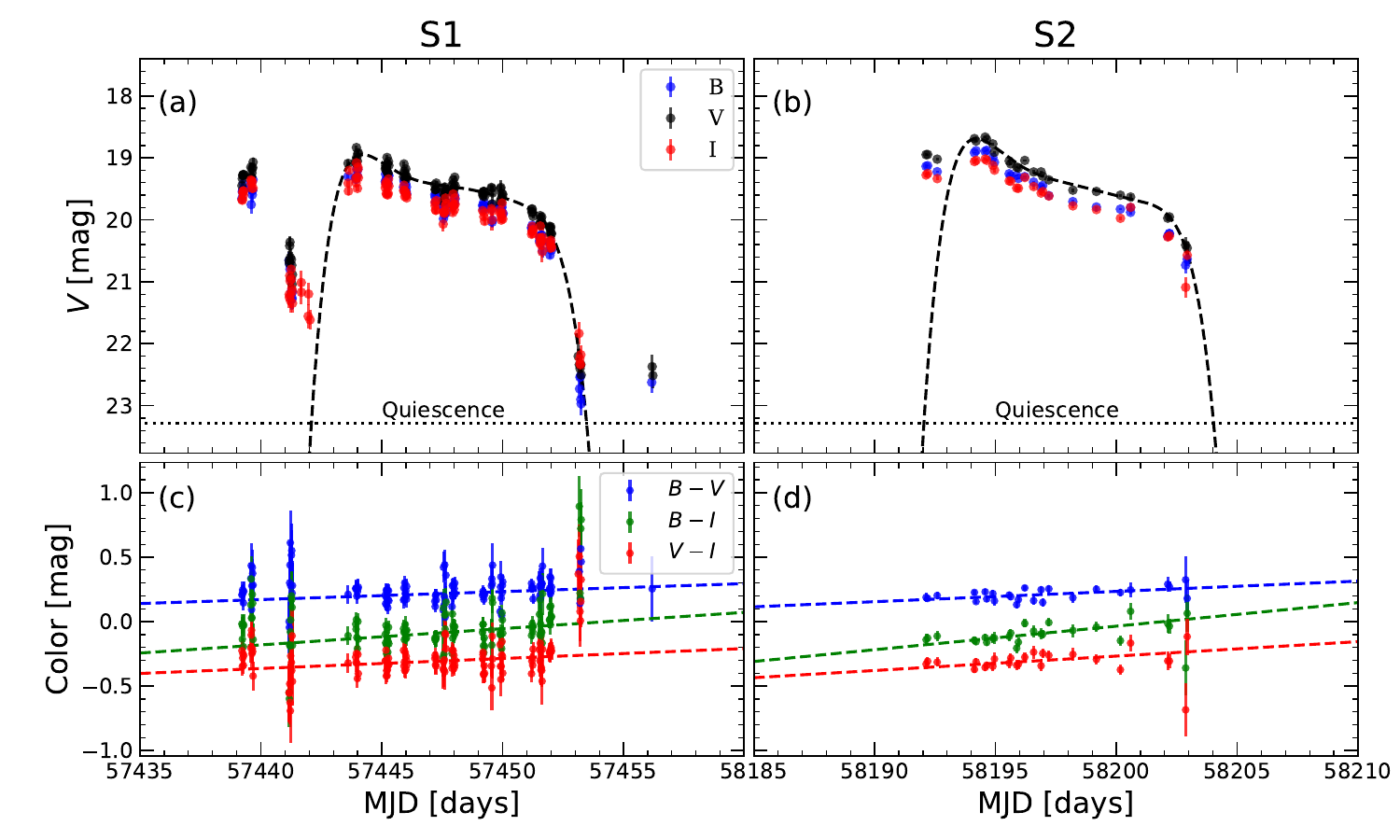}
\caption{Light curves (upper panels) and color evolutions (lower panels)
  of the two SOs of \target: left panels for S1 and right panels for S2.
In panels (a) and (b),
  quiescent phase magnitude ($V\approx 23.29 \pm 0.17$ mag) is plotted
  as dotted lines, and
  sixth-order polynomial fittings for the $V$-band data are shown 
  as black dashed lines.
Dashed lines in panels (c) and (d) are order 1 polynomial fittings
  for each color.
}
\label{fig:solc}
\end{figure*}

Figure~\ref{fig:solc} shows the LCs (upper panels) and 
  color profiles (lower panels) of the two SOs of \target.
In panels (a) and (b),
  $B$ (blue), $V$ (black), and $I$-band (red) observations are shown
  in filled circles, and the sixth-order polynomial fittings for the $V$-bands
  are shown in black dashed lines.
Both SOs are composed of three parts: rising, plateau, and declining parts.
From the polynomial fittings, the peaks are found to be:
  $V = 18.95 \pm 0.03$ mag at MJD $=57444.2197$ for S1 and
  $V = 18.69 \pm 0.05$ mag at MJD $=58194.2067$ for S2
  (larger error for S2 might be due to lack of data at the rising part).
If we consider the time brighter than the quiescent-phase magnitude,
  the SO durations are 11.39 days (S1) and 12.00 days (S2).
The magnitude amplitude between the peak and the quiescent-phase
  is 4.34 mag (S1) and 4.60 mag (S2).
The rising, plateau, and decline rates are
  3.35, 0.12, and 2.05 mag day$^{-1}$, respectively, for S1 and
  3.43, 0.15, and 2.15 mag day$^{-1}$, respectively, for S2.

$B-V$ and $V-I$ color profiles in Figure~\ref{fig:solc} (c) and (d)
  remain almost constant during the SOs with very small slopes of
  0.006 and 0.008 mag day$^{-1}$, respectively, for S1 and
  0.008 and 0.011 mag day$^{-1}$, respectively, for S2.
Only the combined $B-I$ profiles show somewhat larger slopes of
  0.013 (S1) and 0.018 (S2) mag day$^{-1}$.
The mean color values are $B-V \approx 0.23 \pm 0.08$,
  $V-I \approx -0.30 \pm 0.08$, and 
  $B-I \approx -0.06 \pm 0.10$ mag for S1, and
    $B-V \approx 0.21 \pm 0.04$,
    $V-I \approx -0.30 \pm 0.05$, and 
    $B-I \approx -0.10 \pm 0.07$ mag for S2.

%--------------------------------------------------------------------------------
\subsection{Normal Outbursts}         \label{sec:NOsec}

\begin{figure*}   % Fig 4 ---------------------------
\epsscale{1.0}
\plotone{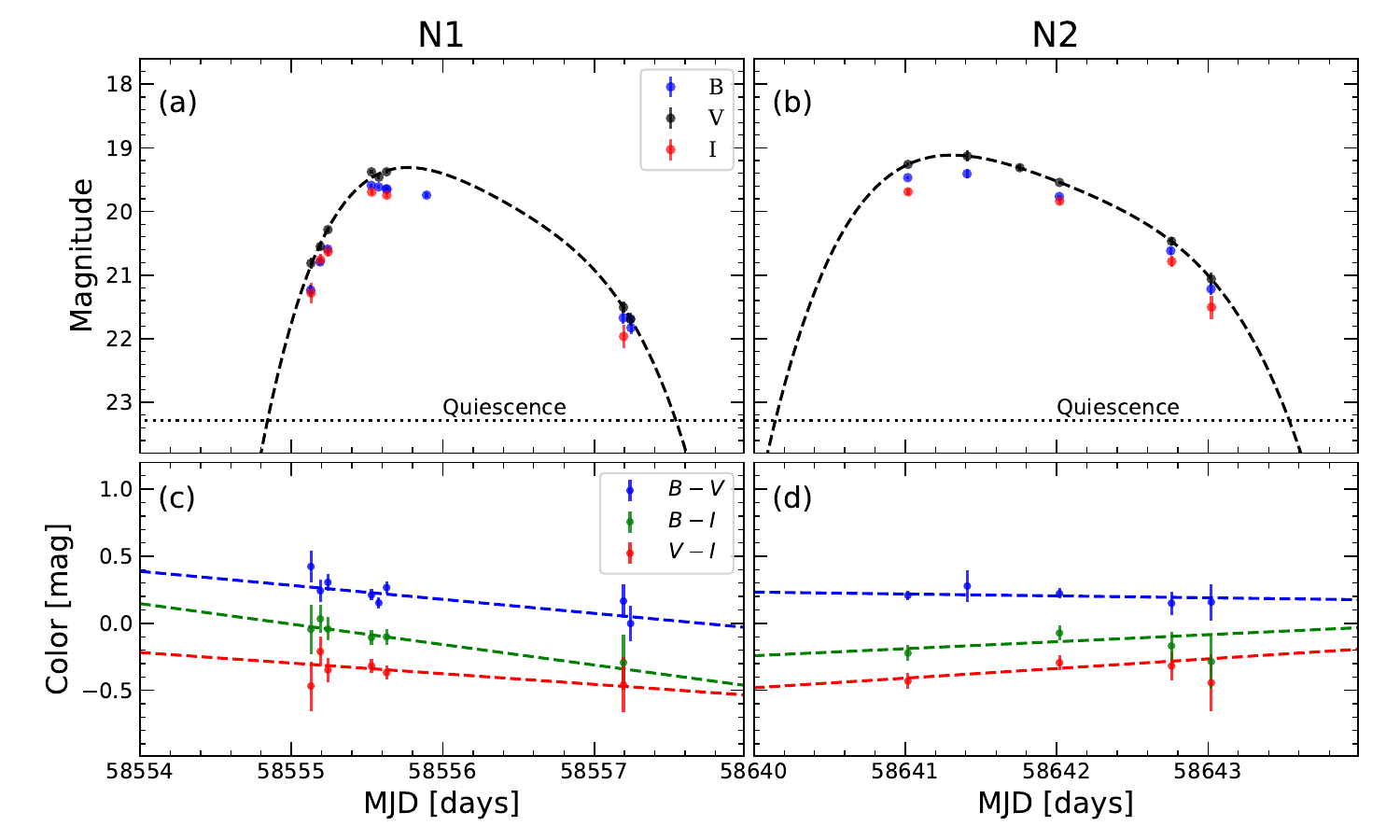}
\caption{Light curves (upper panels) and color evolutions (lower panels) 
  of the two NOs of \target: N1 in left panels and 
  N2 in right panels.
In panels of (a) and (b), $B$, $V$, and $I$-band observation data are
  shown in blue, black, and red filled circles, respectively.
The black dashed lines are the fourth-order polynomial fittings 
  to the $V$-band data.
Black horizontal dotted lines display the $V$-band quiescent phase magnitude
  of $V\approx 23.29 \pm 0.17$ mag.
In panels (c) and (d), $B-V$, $B-I$, and $V-I$ colors are shown 
  in blue, green, and red symbols, and order 1 polynomial fitting 
  results are plotted as dashed lines.
}
\label{fig:nolc}
\end{figure*}

Figure~\ref{fig:nolc} shows the LCs (upper panels) and 
  color profiles (lower panels) for the two NOs of \target.
In the upper panels, $B$ (blue), $V$ (black), and $I$-band (red)
  observations are shown in filled circles, while
  the fourth-order polynomial fittings onto the $V$-band data
  are displayed in black dashed lines.

From the $V$-band polynomial fittings, we get the peak magnitudes of
  $V = 19.37 \pm 0.18$ mag (N1) and
  $V = 19.11 \pm 0.01$ mag (N2) at
  MJD $=58555.7702$ and 58641.3099 days, respectively
  (the large error for N1 might be due to less number of data
  and the scatter of data near the peak).
The magnitude amplitudes are 
  $\Delta V = 3.98$ mag (N1) and $\Delta V = 4.18$ mag (N2) and
  the outburst durations above the quiescent phase magnitude are
  2.68 days (from 58554.8492 to 58557.5319 days) for N1 and 
  3.38 days (from 58640.1487 to 58643.5321 days) for N2.

While the colors of the two NOs still show weak gradient,
  they are somewhat steeper than those of the SOs.
The slopes of N1 colors are $-0.10$ for $B-V$, $-0.08$ for $V-I$, and
  $-0.15$ for $B-I$, and those for N2 are
  $-0.01$ for $B-V$, 0.07 for $V-I$, and 0.05 for $B-I$.
The mean colors are $B-V \approx 0.22$, $V-I \approx -0.36$, and
  $B-I \approx -0.09$ mag for N1, and
  $B-V \approx 0.20$, $V-I \approx -0.37$, and
  $B-I \approx -0.19$ mag for N2.

%--------------------------------------------------------------------------------
\subsection{Period Analysis}         \label{sec:PDMperiod}

\begin{figure*}   % Fig 5 ---------------------------
\epsscale{1.0}
\plotone{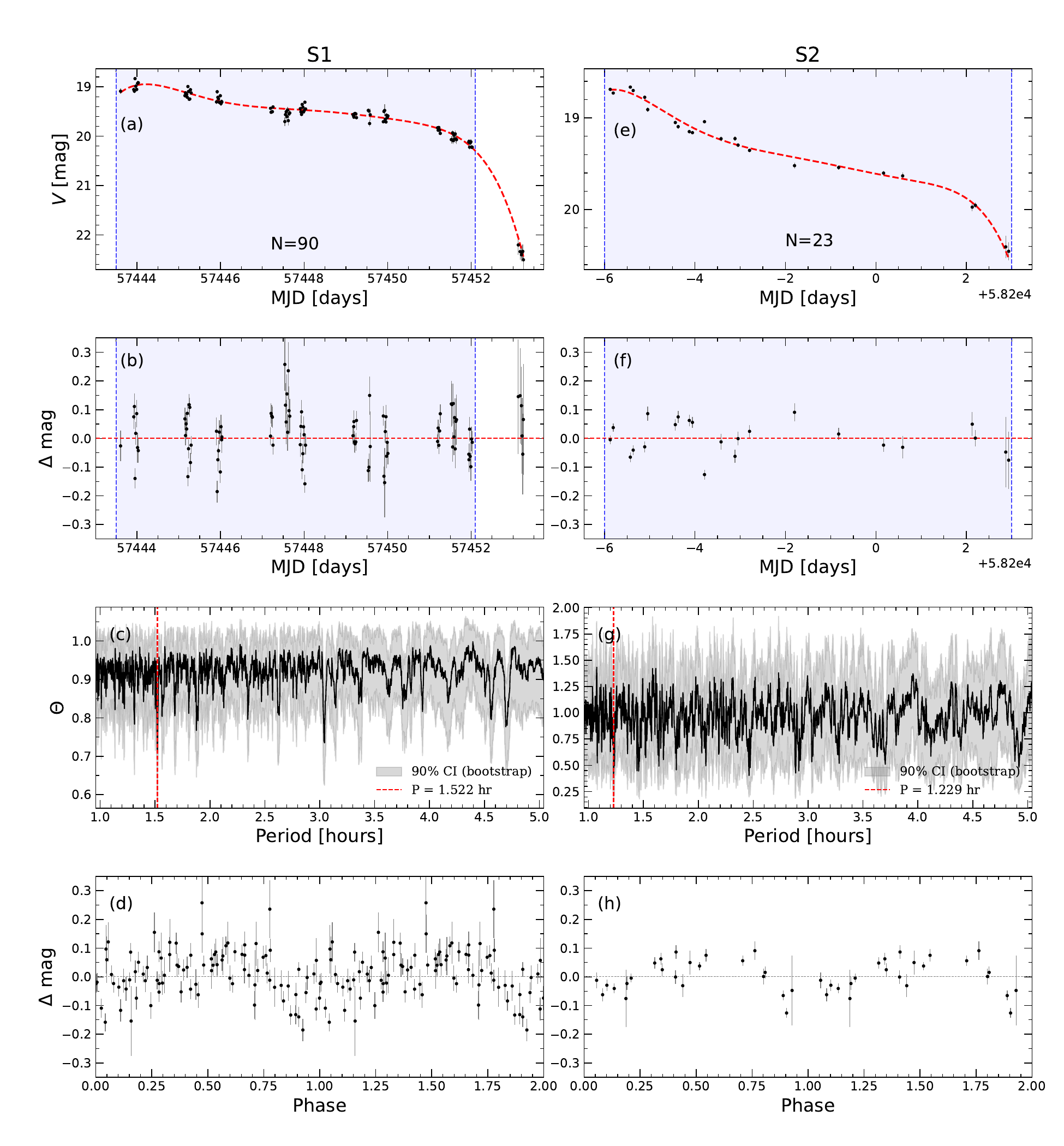}  % 260828_Fig0002_K196_1_1_V0050_PDM.pdf
\caption{Phase dispersion minimization (PDM) results for the $V$-band data of \target\
  for the two SOs: S1 (left panels) and S2 (right panels).
Top panels ((a) and (e)) show the LCs of the SO plateau parts: black filled circles
  are $V$-band observation data and red dashed lines
  are best 6th-degree polynomial fits to be subtracted from the observation data.
The blue shaded areas display the regions used for the PDM analyses.
Second panels ((b) and (f)) show the polynomial fit subtracted $\Delta$ mag plots.
Third panels ((c) and (g)) are the PDM results: black solid lines and 
  the shaded areas indicate median $\Theta$ and 90\% confidence interval
  from 300 bootstrap PDM, respectively.
The red vertical dashed lines demonstrate the minimum values of $\Theta$
  for each SOs.
Lowest panels ((d) and (h)) are phase-fold profiles.
}
\label{fig:PDMV}
\end{figure*}

We estimate the superhump periods using the phase dispersion minimization 
  (PDM: \citealt{Stellingwerf78}) 
  and Lomb–Scargle (LS) method \citep{lomb76,scargle82}\footnote{Astropy.stats.LombScargle 
  is implemented
  \citep[\url{http://docs.astropy.org/en/stable/stats/lombscargle.html};][]{Ast13,Ast18}},
  both of which are commonly used 
  to detect and characterize periodic signals in unevenly sampled time-series data
  \citep{LinnellNemec85,SchwarzenbergCzerny97,Tampo26, % PDM
  lee19,lee22,lee24,  %LS
  kim26a,kim26b}. % PDM
Unlike the LS method, which assumes a sinusoidal variation,
  the PDM method determines the most likely period 
  directly from the observed light curve 
  without assuming a specific functional form for the periodic variation.
Figure~\ref{fig:PDMV} illustrates the procedure used to determine 
  the $V$-band superhump period with the PDM method. 
Figure~\ref{fig:PDMV} (a) and (e) show the 6th-order polynomial fits 
  to the $V$-band LCs obtained during the plateau phase of \target. 
After subtracting the best polynomial fits from the original LCs, 
  we obtain the residual LCs shown in Figure~\ref{fig:PDMV} (b) and (f). 
We then apply the PDM analysis to these residual LCs, and 
  the resulting periodograms are presented in Figure~\ref{fig:PDMV} (c) and (g). 
The strongest minima in the PDM periodograms, corresponding to the superhump period,
  yield the periods of 
  $P_{\rm sh} \approx 1.522$ hr for S1 and 
  $P_{\rm sh} \approx 1.229$ hr for S2. 
The bottom panels ((d) and (h)) show the corresponding phase-folded LCs 
  constructed using the derived superhump periods above.
The LS method applied to the $V$-band data of S1 and S2 yield 
  $P_{\rm sh} \approx 1.521$ hr for S1 and
  $P_{\rm sh} \approx 1.231$ hr for S2.

In order to increase the statistical reliability, % signal-to-noise ratio and get better cadence
  we combine the $B$, $V$, and $I$-band residual magnitudes 
  after subtracting the 6th-order polynomial fits.
For these $BVI$-combined data, the PDM method gives
  $P_{\rm sh} \approx 1.522$ hr for S1 and
  $P_{\rm sh} \approx 1.898$ hr for S2, and
  the LS method produces
  $P_{\rm sh} \approx 1.520$ hr for S1 and
  $P_{\rm sh} \approx 1.460$ hr for S2.

Compared to Figure~\ref{fig:PDMV} (a), Figure~\ref{fig:PDMV} (b) shows
  poorer quality of cadence and number of data during the plateau
  to get good enough superhump period for \target.
While the combination of four methods described above yield
  a consistent period for S1, the results for S2 vary a lot
  from 1.23 to 1.90 hr. %for the superhump period of \target.
For the superhump period of \target, therefore, we adopt
  the results for S1: 
  $P_{\rm sh} \approx 1.52 ^{+0.04}_{-0.01}$ hr
  ($=91.2$ minutes $=0.06333$ days).
The errors for the superhump period are obtained
  using data within 10\% of the measured peak period. %of 1.52 hr.
We performed 300 bootstrap iterations and 
  calculated the 16th–84th percentile range around 1.52 hr.

%==============================================================================
\section{Discussion}\label{sec:discuss}  % Sec 4

\subsection{Precursor Outburst}         \label{sec:precursor}
\begin{figure*}   % Fig 6 ---------------------------
\epsscale{0.9}
\plotone{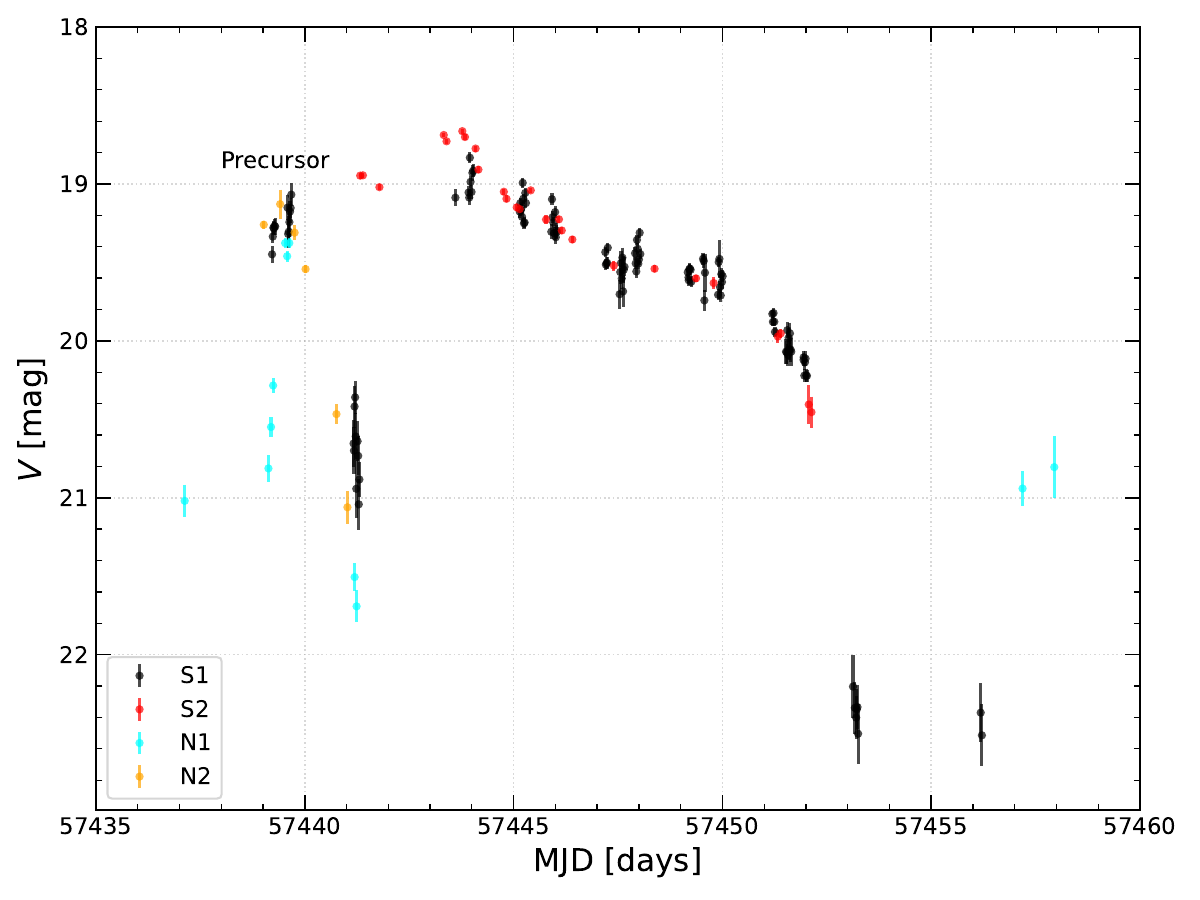}
\caption{Overplot of two SOs and two NOs of \target.
Abscissa is set for S1, while S2, N1, and N2 data are shifted 
  by $-750.8$, $-1116$, and $-1202$ days, respectively.
}
\label{fig:overlap}
\end{figure*}

Figure~\ref{fig:overlap} shows the overplot of the two SOs
  (S1 (black) and S2 (red)) and two NOs (N1 (cyan) and N2 (orange)), 
  which shows a clear precursor (a short shoulder immediately preceding
  the SOs) in S1 before MJD $=57440$ days.
Using the Kepler light curve of V1504 Cygni, \citet{Osaki13_50} studied
  the SOs and found that the superhump first appears near the peak %maximum
  of the precursor.
The superhump amplitude gradually increases from the precursor 
  to the main SO, and
  the epoch of maximum superhump amplitude agrees fairly well
  with that of the peak brightness of the SO.
This suggests that the SO is triggered by tidal instability,
  and is a supporting evidence of the TTI model 
  \citep{Osaki89,Kato12}. % thermal-tidal instability
For this reason, the precursor is known to be an NO triggering an SO 
  and the superhump (the tidal instability) is likely the cause of the SO
  \citep{Polidan87,Osaki03AA,Schreiber04,Osaki05_PJAB,Osaki13_50,
  kato16,Kato19_L4,Dai21,Szegedi22}.
  %of the precursor that is a supporting evidence of 
  %the thermal-tidal instability model \citep{Osaki89,Kato12}. %TTI

\citet{Tampo23} compared three SOs of the WZ Sge-type DN V627 Pegasi %(Peg) 
  observed in 2010, 2014, and 2021, and showed that the 2010 SO 
  exhibited the characteristics of a WZ Sge-type SO
  (showing `early superhump' phase),
  whereas the SOs observed in 2014 and 2021 showed those of SU UMa-type SOs.
In particular, their 2021 observations revealed a clear precursor,
  which had never been observed in WZ Sge-type DNe and thus 
  provide strong evidence for its SU UMa-type nature.
This demonstrates that even within a single system, 
  a precursor may appear in some SOs but not in others
  \citep{Uemura05}.

For the possible ER UMA-type (a sub-group of SU UMa
  with high and active mass-transfer rate) DN KSP-OT-201701a, 
  \citet{lee22} have shown three SOs and six NOs during $\sim 2.6$ yr.
All the three SOs in their LC show precursors lasting $\sim 3$ days
  and $2-5$ days before the peak brightness (see their figure 1).
Though the overall shapes of the precursors are similar,
  the transitions from the precursor to the SO differ 
  for each precursor.

If the three S2 points before the peak magnitude in Figure~\ref{fig:overlap}
  are increasing in brightness 
  with time, they could be considered part of the rising phase. 
However, since they become fainter with time, they may instead 
  indicate a precursor phenomenon, although this interpretation 
  remains uncertain due to the limited available data.
As shown in Figure~\ref{fig:overlap}, 
  the LC morphology of the precursor is very similar to 
  that of N1 and N2 in terms of peak brightness, width, and other characteristics, supporting the claim that 
  the precursor is an NO responsible for triggering the SO.

%------------------------------------------------------------------------------
\subsection{Outburst Parameters}         \label{sec:type}
  % SU UMa-type Identity

\begin{figure*}   % Fig 7 ---------------------------
\epsscale{0.9}
\plotone{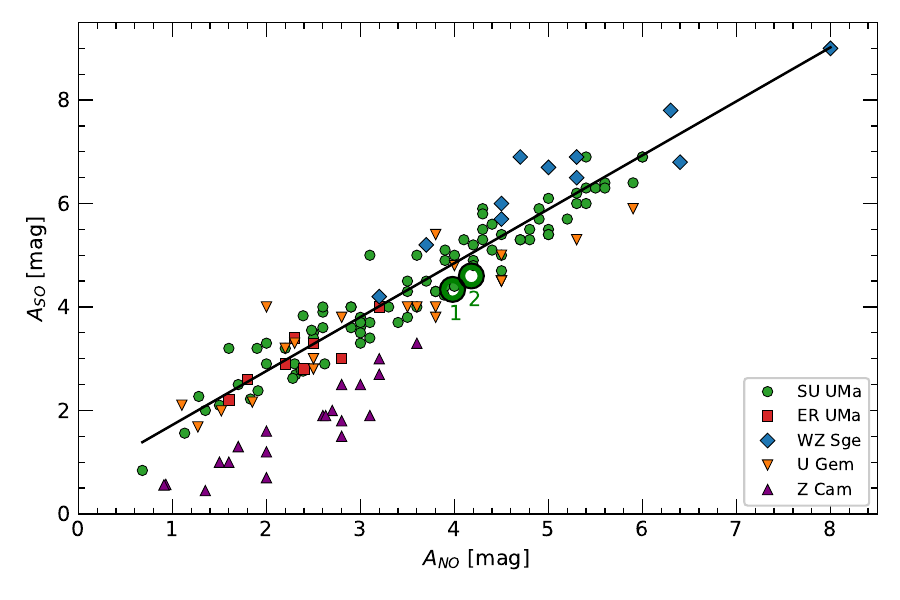} %Parameter_relations.png}
\caption{Correlations between the %physical parameters of DNe:
  outburst property parameters of DNe:
  normal outburst amplitude $A_{\rm NO}$ in abscissa and
  superoutburst amplitude $A_{\rm SO}$ in ordinate.
Data are from \citet{Otula16}, while large green circles
  labeled `1' and `2' denote N1-S1 and N2-S2 combinations,
  respectively, for \target.
Black solid line is a linear fit for DNe of SU UMa, ER UMa, and
  WZ Sge types. (See text for details.)
}
\label{fig:OtulaAmp}
\end{figure*}

Figure~\ref{fig:OtulaAmp} shows the distribution of 
  NO amplitude $A_{\rm NO}$ (abscissa)
  versus SO amplitude $A_{\rm SO}$ (ordinate),
  with data from \citet{Otula16} and 
  N1-S1 (labeled as `1') and N2-S2 (labeled as `2') combinations
  for \target.
While the NO amplitudes for N1 and N2 are in the middle of SU UMa-types
  (green circles),
  the SO amplitudes for S1 and S2 are at the lower boundary
  of the SU UMa-type DNe.
%The distribution of 
Z Cam-type DNe are located much below the trend for other types
  in Figure~\ref{fig:OtulaAmp}, while they mostly sit in the middle
  following the fit line in Figure 13 of \citet{Otula16}.
We doubt there be some errors in their Table 2.

\begin{figure*}   % Fig 8 ---------------------------
\epsscale{0.9}
\plotone{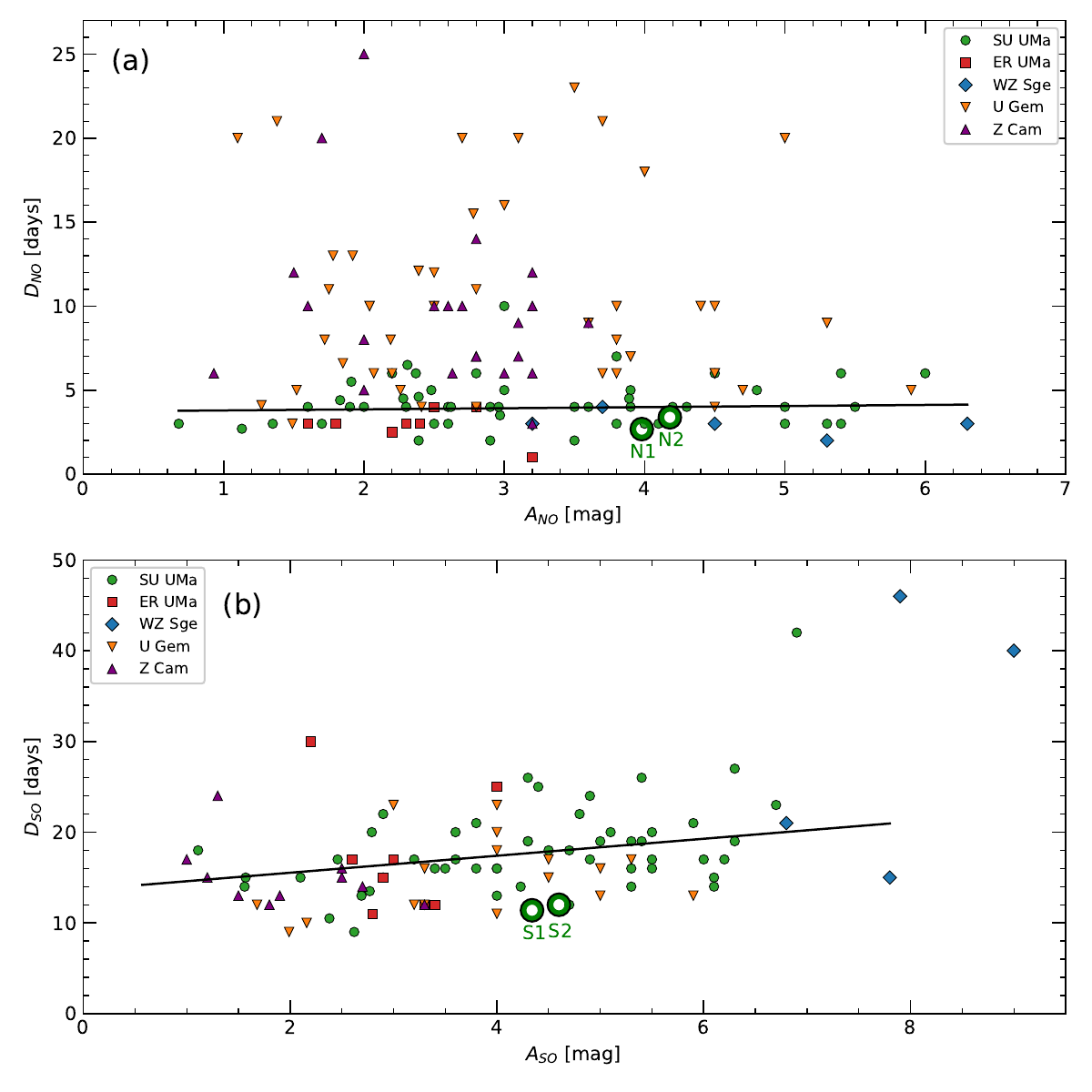}
\caption{Correlations between the outburst amplitudes (abscissa)
  and durations (ordinate) of DNe: (a) for NOs and (b) for SOs.
$A_{\rm NO}$ and $A_{\rm SO}$ are amplitudes of NOs and SOs, respectively, and
  $D_{\rm NO}$ and $D_{\rm SO}$ are durations of NOs and SOs, respectively.
Data are from \citet{Otula16}, while large green circles 
  are for \target.
Black solid lines are linear fits for DNe of SU UMa, ER UMa, and
  WZ Sge types. (See text for details.)
}
\label{fig:OtulaDur}
\end{figure*}

Figure~\ref{fig:OtulaDur} shows the distributions of amplitudes (abscissa)
  and durations (ordinate) of NOs (panel (a)) and SOs (panel (b)).
Although we plotted linear fitting results as solid lines,
  both NOs and SOs seem to show no or weak correlations:
  %As \citet{Otula16} explains, there seems to be no correlation 
  correlation coefficients for the NOs $r=0.06$ and 
  $r=0.26$ for the SOs.
  %while panel (b) shows a weak relation %of 
  %$D_{\rm SO} = 2.09(0.43) \cdot A_{\rm SO} + 9.7(2.0)$ 
  %with correlation coefficient r=0.52.
  %and the standard error of the estimate of 5.7
  %(values in parentheses are errors in the last digits).
Figure~\ref{fig:OtulaDur} (a) shows that the two NOs of \target\
  are well among the distribution of SU UMa-type DNe, %while
  though they are still close to the shorter boundary
  of $D_{\rm NO}$ values.
On the other hand, Figure~\ref{fig:OtulaDur} (b) reveals that 
  the SO durations of \target\ are almost at the lowest boundary %values
  of the SU UMa-type DNe, while the amplitudes are
  quite in the middle.
  %(amplitude : at the lowest end, Figure 8)

\citet{Osaki05_PJAB} have reviewed the SOs of DNe % explaine
  using four disk instability models, among which
  Types A and B are SO cases with and without precursor,
  respectively.
In Type A with the precursor outbursts, an NO expands the disk
  beyond the 3 : 1 resonance radius, which rapidly increases
  the tidal instability.
An eccentric disk develops during the precursor outburst,
  and enhanced tidal dissipation keeps the outer disk hot,
  producing an SO.
Superhumps develop together with the rise toward the SO peak
  (see figure 4 of \citealt{Osaki05_PJAB}).
On the other hand, Type B (their figure 5) occurs 
  when the disk reaches the 3 : 1 resonance, but 
  the eccentric instability does not grow quickly enough.
So the disk begins to shrink again, and the NO fails to 
  trigger an SO.
After subsequent NOs, the disk gradually expands farther
  to reach the tidal truncation radius eventually.
Then, an SO develops  with long-lived viscous plateau phase,
  during which the eccentric disk finally develops and
  the superhump appear (a few days after the SO peak).

The precursor outburst may modify the radial surface-density
  distribution and disk radius, thereby altering
  the initial conditions for the subsequent SO.
To better understand why SOs with precursors have
  shorter durations ($D_{\rm SO}$ in Figure~\ref{fig:OtulaDur} (b)) 
  despite having rather typical SO amplitudes ($A_{\rm SO}$),
  model calculations incorporating 
  the radial surface-density distribution and disk radius may be needed
  in the future (see, e.g., \citealt{Truss00,Truss01,Lasota01}).

%==============================================================================
\section{Summary and Conclusions}\label{sec:summary}  % Sec 5

In this paper, we report the discovery and observations of
  \target, an SU UMa-type dwarf nova,
  based on $BVI$-band photometry obtained from
  the KMTNet Supernova Program (KSP).
We summarize our results and conclusions as follows.
\begin{itemize}
\item
From KSP observations from 2016 to 2019,
  \target\ is identified as an SU UMa-type DN with
  two superoutbursts (SOs: S1 and S2) and two normal outbursts 
  (NOs: N1 and N2).
\item
The peak brightnesses are found to be 
  $V = 18.95$ mag (S1), $V = 18.69$ mag (S2) and 
  $V=19.37$ mag (N1), $V=19.11$ mag (N2).
\item 
The magnitude amplitudes for the SOs are 
  4.34 mag (S1) and 4.60 mag (S2), and those for the NOs are
  3.98 mag (N1) and 4.18 mag (N2).
\item 
The durations of the SOs are 11.39 days (S1) and 12.00 days (S2),
  and those for the NOs are 2.68 days (N1) and 3.38 days (N2).
\item
The phase dispersion minimization analysis on superhump of S1
  affords the superhump period of 1.52 hr ($=91.2$ minutes $=0.06333$ days),
  while the data for S2 are not sufficient enough
  to extract useful period.
\item
In S1, clear evidence of a precursor is identified before the SO peak,
  supporting the hypothesis that the precursor phenomenon is 
  an NO triggering an SO and the superhump.
In contrast, for S2, only a possible indication of the precursor phenomenon
  is observed, based on limited data.

\end{itemize}

%==============================================================================
\begin{acknowledgments}
We thank the referee for the very helpful comments that improved the manuscript.
This research has made use of the KMTNet system 
  operated by the Korea Astronomy and Space Science Institute (KASI) 
  at three host sites of CTIO in Chile, SAAO in South Africa, and SSO in Australia.
Data transfer from the host site to KASI was supported by
  the Korea Research Environment Open NETwork (KREONET).

This research was supported by the Korea Astronomy and Space Science Institute 
  under the R\&D program (Project No. 2026-1-860-00) 
  supervised by the Ministry of Science and ICT.
\end{acknowledgments}

%\software{KCORRECT v4.2 \citep{Bla07}}

\bibliography{draft}
\bibliographystyle{aasjournalv7}

\end{document}